\documentclass[prl,twocolumn,preprintnumbers,amsmath,amssymb]{revtex4}
\usepackage{graphicx}
\usepackage{dcolumn}
\usepackage{amsmath}
\usepackage{amssymb}
\usepackage{epsfig}

\makeatletter
\def\btt#1{\texttt{\@backslashchar#1}}
\DeclareRobustCommand\bblash{\btt{\@backslashchar}}
\makeatother

\begin{document}

\preprint{}
\title{Comment on ``Quantum waveguide array generator for
performing Fourier transforms: Alternate route to quantum computing''
[APL {\bf 79}, 2823 (2001)]}

\author{Daniel A. Lidar}
\email{dlidar@chem.utoronto.ca}

\affiliation{Chemical Physics Theory Group, University of Toronto, 
80 St. George Street, Toronto, Ontario M5S 3H6, Canada }

\maketitle

In their letter \cite{Akis:01} Akis \& Ferry propose a quantum
waveguide array approach for performing quantum Fourier
transforms (QFTs). The waveguide produces $2^n$ waves at its output
with controllable relative phases; $n$ is the number of binary splits of the
input wave. The interference pattern from these waves is recorded
and implements a Fourier transform. The authors claim that their waveguide approach is ``a
more practical means'' and an alternative to the ``qubit paradigm that
currently dominates the field of quantum computing'' (double quotation
marks are direct quotes from \cite{Akis:01}). The main result claimed
by the authors is an implementation of the QFT
that is as efficient as that obtained using the standard paradigm. In
their conclusions they say:
``... it is unclear whether the promised speedup in certain
computations arises from the quantum nature of the systems or from the
highly parallel analog processing that is provided by the array of
qubits. We have argued that it is the latter that is important, and
that equal speedup is available using analog processing arrays whose
operation is based on general wave principles.''

The arguments leading to this conclusion are unfortunately based on
an incorrect assumption: that {\em interference} is sufficient to
obtain a quantum speedup. The essence of the
waveguide approach is quantum interference. Indeed, the authors
claim: ``Given that quantum mechanics is primarily a wave mechanics
concept, these examples based on electromagnetic and acoustic waves
suggest that there should be a more natural approach to quantum signal
processing than that found in the existing quantum computing
literature.''

It is by now well appreciated that
the exponential speedup offered by quantum computers
in computing the QFT {\em is impossible without entanglement}
\cite{comment}. Detailed discussions of this issue exist in the
literature, e.g., \cite{Ekert:98}. Most recently, Jozsa and Linden
proved that for any quantum algorithm operating on pure states, the
presence of multi-partite entanglement, with a number of parties that
increases unboundedly with input size, is necessary if the quantum
algorithm is to provide an exponential speedup over classical
computation (Theorem 1, \cite{Jozsa:02}). Entanglement is a property that
depends on the existence of a {\em tensor product} Hilbert
space. This implies that it is possible to {\em efficiently}
(i.e., with resources that scale polynomially in the number of qubits)
construct {\em local} (e.g., single- and two-qubit) operators, even though such operators are
represented by exponentially large matrices. It is further understood that
approaches to quantum computing that rely on interference alone,
always incur some form of exponential overhead (in energy, resolution,
or number of building blocks of the quantum circuit)
\cite{Meyer:00a,Meyer:00}.

The waveguide approach of Akis \& Ferry is no different: by relying on
interference, without entanglement, the authors have eliminated a key
ingredient of the quantum speedup. Their proposed devise is not
equivalent to the standard qubit paradigm of quantum computing because
it does not support a tensor-product Hilbert space. It is a
multi-level quantum system, which has computational power equivalent to an experiment in
classical wave mechanics. The exponential overhead they incur is in
the size of their waveguide, as is immediately evident from Fig.~1(b)
in their paper. Their waveguide has the shape of a binary tree; the
distance between its nodes (the radiating elements) cannot be made
arbitrarily small. Hence the overall size of the device must grow
exponentially. This can certainly not qualify as a valid quantum computer.

Support from the DARPA-QuIST program (managed by AFOSR under agreement
No. F49620-01-1-0468) is gratefully acknowledged.


\begin{thebibliography}{1}
\bibitem{comment}
Note that the exponential speedup of the QFT on a quantum computer is still
  unproven, i.e., a classical (probabilistic) algorithm that is as
fast as the quantum one may still be discovered, although this seems
unlikely.
\end{thebibliography}
\end{document}